\documentclass[10pt,conference]{IEEEtran}
\IEEEoverridecommandlockouts
\pdfoutput=1 
\usepackage{cite}
\usepackage{amsmath,amssymb,amsfonts}
\usepackage{algorithm}
\usepackage[noend]{algpseudocode}   
\usepackage{caption}
\usepackage{graphicx}
\usepackage{textcomp}
\usepackage{multirow}
\usepackage{dcolumn}
\usepackage{bm}
\usepackage{braket}
\usepackage{booktabs}
\usepackage{tikz}
\usetikzlibrary{shapes.geometric, arrows, arrows.meta, positioning, calc}
\usepackage[compat=0.3]{yquant}
\usepackage{hyperref}

\hypersetup{
    colorlinks = true,
    citecolor = magenta,
    linkcolor = purple
}
\def\BibTeX{{\rm B\kern-.05em{\sc i\kern-.025em b}\kern-.08em
    T\kern-.1667em\lower.7ex\hbox{E}\kern-.125emX}}

\tikzstyle{startstop} = [rectangle, rounded corners, minimum width=3cm, minimum height=1cm,text centered, draw=black, fill=red!30]

\begin{document}
\bstctlcite{IEEEexample:BSTcontrol}

\title{A Troubleshooting Framework for Trapping Ions
\thanks{This work is supported by JST Moonshot R\&D Grant No. JPMJMS226C.}}

\author{
\IEEEauthorblockN{
Monet Tokuyama Friedrich\IEEEauthorrefmark{2},
Alto Osada\IEEEauthorrefmark{4},
Rodney Van Meter\IEEEauthorrefmark{1}\IEEEauthorrefmark{2},
and Shota Nagayama\IEEEauthorrefmark{3}\IEEEauthorrefmark{5}
}\\
\IEEEauthorblockA{\IEEEauthorrefmark{1}\textit{Faculty of Environment and Information Studies, Keio University Shonan Fujisawa Campus, Kanagawa, Japan}}
\IEEEauthorblockA{\IEEEauthorrefmark{2}\textit{Graduate School of Media and Governance, Keio University Shonan Fujisawa Campus, Kanagawa, Japan}}
\IEEEauthorblockA{\IEEEauthorrefmark{4}\textit{QIQB, Center for Quantum Information and Quantum Biology, Osaka University, Osaka, Japan}}
\IEEEauthorblockA{\IEEEauthorrefmark{3}\textit{Keio University Graduate School of Media Design,  Hiyoshi Campus, Kanagawa, Japan}}
\IEEEauthorblockA{\IEEEauthorrefmark{5}\textit{mercari R4D, Mercari, Inc., Tokyo, Japan}}\\
\{bob, rdv\}@sfc.wide.ad.jp\\ osada.alto.qiqb@osaka-u.ac.jp \\ shota@qitf.org
}

\thispagestyle{plain}
\pagestyle{plain}
\maketitle

\begin{abstract}
Practical knowledge about troubleshooting and error handling in trapped-ion systems remains largely undocumented and held within individual labs, creating a barrier to cross-disciplinary collaboration towards engineering scalable systems. This paper presents a structured troubleshooting framework for trapping ions, developed through hands-on experience in the lab. The framework categorizes standard failure modes across subsystems—vacuum, electronics, optics, and imaging—and organizes them into a modular decision-tree structure. Each troubleshooting action is annotated with estimates of cost and operational risk inspired by the principles of Failure Mode and Effects Analysis (FMEA). By categorizing failure modes and their associated costs, this work bridges the gap between physicists and engineers, enabling collaborative system design and setting the stage for turning experimental setups into robust, reproducible products. This framework provides a structured basis for future error-handled ion-trap systems, designed with diagnostics and maintainability in mind—for use in distributed quantum computing and networks.
\end{abstract}


\section{Introduction}
\label{sec:introduction}
Realizing large-scale quantum computers will require distributed architectures~\cite{meter2006} that connect multiple quantum processors through a quantum network~\cite{rfc9340}. Distributed quantum computing has gained significant attention through recent works~\cite{q-fly}. Trapped-ion systems are a leading candidate~\cite{science.1231298, Lekitsch2017Blueprint} among the potential platforms due to their long coherence times, well-controlled qubits, and potential for photonic interconnect~\cite{akhtar2023high}. However, constructing and operating these systems remains an engineering and operational challenge, especially on a large scale. Their construction and performance depend on precise calibration and troubleshooting, skills that are often undocumented and developed within labs.

The development of ion trap systems is at a critical bottleneck. Even though large numbers of trapped-ion systems are necessary to utilize the strength of such technologies effectively, it is still a difficult system to control. While individual groups/labs may succeed in trapping ions and performing gate operations and interconnects across distances~\cite{PhysRevLett.130.050803}, reproducing and scaling this is still a challenge, an engineering challenge. Scaling such technologies requires robust and reliable designs alongside standardized methods for identifying and correcting faults that may occur. Yet, no widely accepted framework or any open-source platform for practical knowledge exists for troubleshooting trapped-ion systems in a systematic and reproducible way.

This work proposes a structured troubleshooting framework developed through hands-on experience during the construction and debugging of a trapped-ion quantum node. The framework identifies recurring failure modes across the core subsystems— vacuum, electronics, optics, and imaging— and organizes them into a modular, decision-tree-based structure. This approach helps identify the root cause of failure when we are unable to confirm successful trapping while reflecting the step-by-step nature of the whole ion-trapping process.

To evaluate and prioritize the diagnostic steps, we adopt a methodology inspired by Failure Mode and Effects Analysis (FMEA)~\cite{NASA_1966, sharma2018fmea}, a well-established tool in system and reliability engineering. FMEA is an evaluation method that categorizes failure modes and then evaluates each of the costs associated with those failures. We have already categorized and described the relation between different failure modes in ion trap systems through the troubleshooting framework proposed in this paper. While FMEA is typically applied quantitatively using statistical fault data, such an approach is currently infeasible in quantum hardware due to the low volume of reproducible systems, especially within one lab. Instead, we propose a qualitative FMEA-like evaluation based on experimental insight, assigning estimated costs such as time cost, misalignment risk, and operational impact to each action in the troubleshooting framework.
The goal is not only to help researchers debug ion trap systems more effectively but to lay the groundwork for standardizing knowledge across labs and fields, spanning from the physics to the engineering community. By working with FMEA principles, our framework creates a structure that can eventually be used with real-world fault data and integrated into software-assisted or automated debugging tools. This work is the first step towards engineering trapped-ion systems that are not only in-lab quantum devices but robust, maintainable, and scalable products.

\section{Overview of Trapped-Ion Systems}
\label{sec:basic_hardware}
The development of ion traps dates back to the mid-20th century. Wolfgang Paul and Hans Dehmelt were pioneers in this field, leading to significant advancements that earned them the Nobel Prize in Physics in 1989~\cite{Paul_Nobel}. It was in the 1950s that Paul developed the quadrupole ion trap, now commonly known as the \textit{Paul Trap}, which revolutionized ion trapping techniques. In 1995, Cirac and Zoller proposed the usage of ion traps for quantum computing~\cite{IonTrap_QC1}, and by the early 2000s, ion traps were recognized for their potential in quantum computing and networking.

\begin{figure}[h!]
    \centering
        \centering
        \includegraphics[width=0.25\textwidth]{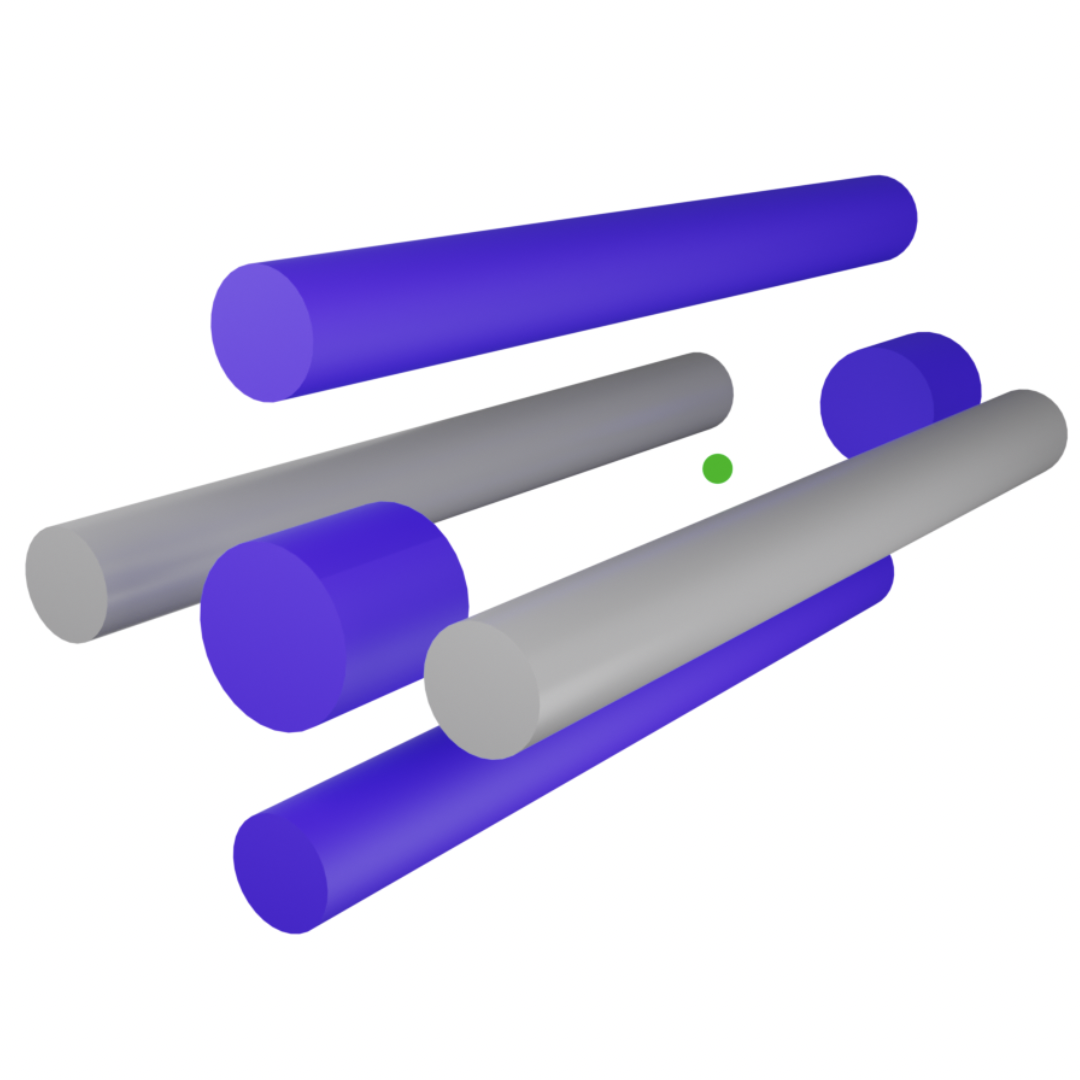}
        \caption{Schematic diagram of a linear trap; purple and silver describing the DC and RF electrodes respectively.}
        \label{fig:linear_trap}
\end{figure}

\begin{figure}[h!]
        \centering
        \includegraphics[width=0.3\textwidth]{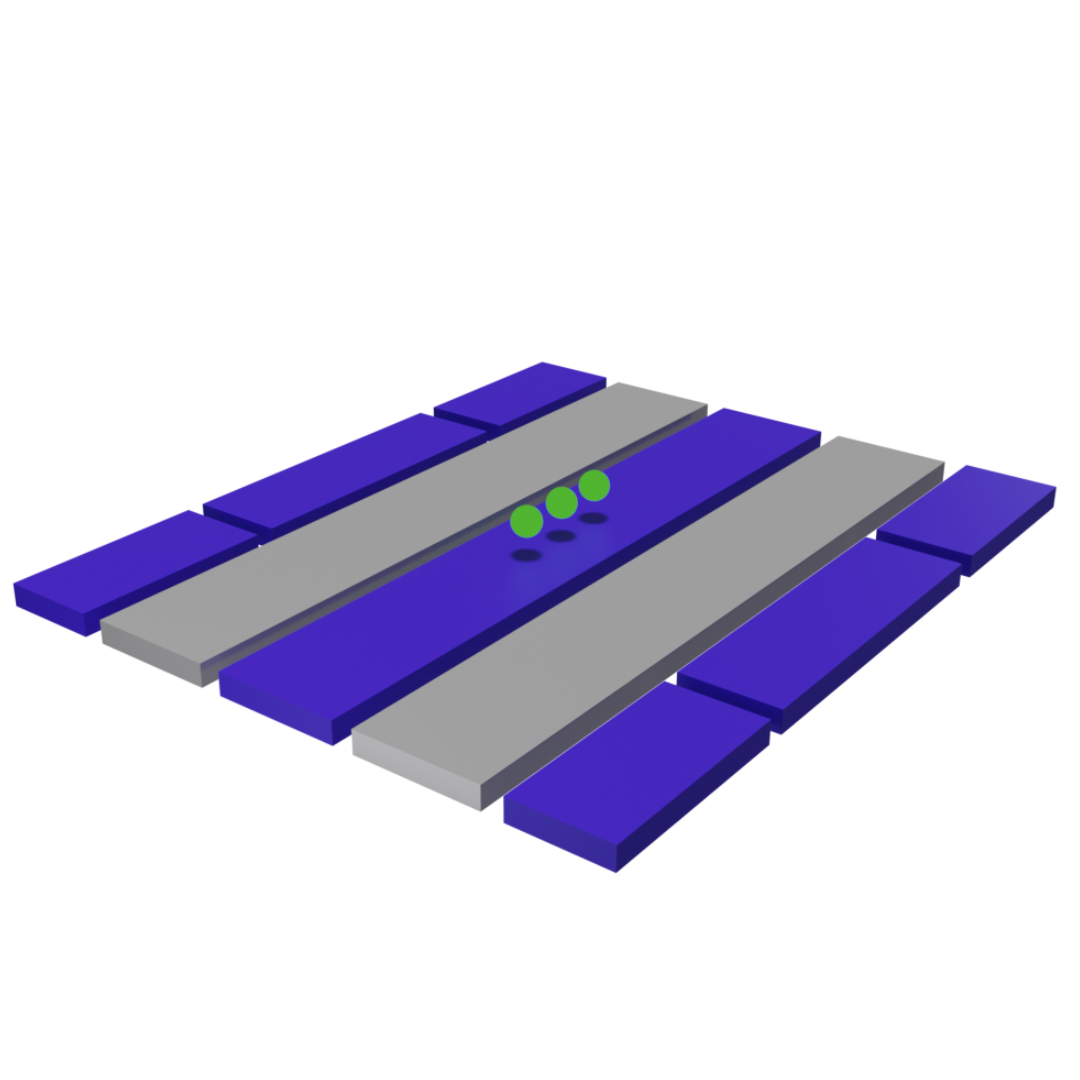}
        \caption{Schematic diagram of a surface trap.}
        \label{fig:surface_trap}
\end{figure}

However, ion traps remain a difficult technology to scale due to the underlining physical principles within the system. Due to these physical constraints, early designs of ion traps were bulky and not suitable for accommodating large numbers of ions, which are used as qubits.
Researchers worked on improving trap designs, such as linear ion traps (Fig~\ref{fig:linear_trap}) and microfabricated surface-electrode traps (Fig~\ref{fig:surface_trap}), allowing potential scalability~\cite{qccd_proposal},~\cite{PhysRevLett.96.253003}. Despite these advances, scaling the number of qubits remains a challenge for ion trap quantum computers due to the following reasons:

\vspace{1em}\begin{enumerate}
\setlength\itemsep{1em}
    \item The number of ions manageable in a single trap is limited (100 at most)
    
    \item There is technical uncertainty and overhead for increasing and connecting the number of traps in a chamber
    
    \item There is a technical overhead for connecting the traps over different chambers, the photonic interconnection
\end{enumerate}

\vspace{1em}Current research focuses on distributed quantum computing and quantum networking, using ion traps as quantum nodes. These approaches may help overcome scalability issues and shift the game towards achieving fault-tolerant quantum computing~\cite{Benhelm_2008}. To construct several nodes that satisfy the qubit count necessary for performing meaningful quantum operations, we will need a large number of trapped-ion systems. To address this engineering challenge of mass-producing technically complex and high-maintenance systems, even for small scales, we need better-structured engineering procedures and reliable, reproducible products.

\subsection{Principles of Ion Trapping}
The underlining principles of ion trapping include the charge of ions and how, due to the charge, ions can be confined in electromagnetic potentials.
Such potentials —often called the trapping potential— are typically expressed as a quadrupolar potential that satisfies Laplace's equation~\cite{RevModPhys.75.281}.

\begin{multline}
\Phi(x, y, z, t) = \frac{U}{2}(ax^2 + by^2 + cz^2) \\
+ \frac{\tilde{U}}{2}\cos(\Omega_{\text{rf}} t)(a'x^2 + b'y^2 + c'z^2)
\end{multline}

\begin{figure}[H]
    \centering
    \includegraphics[width=0.35\textwidth]{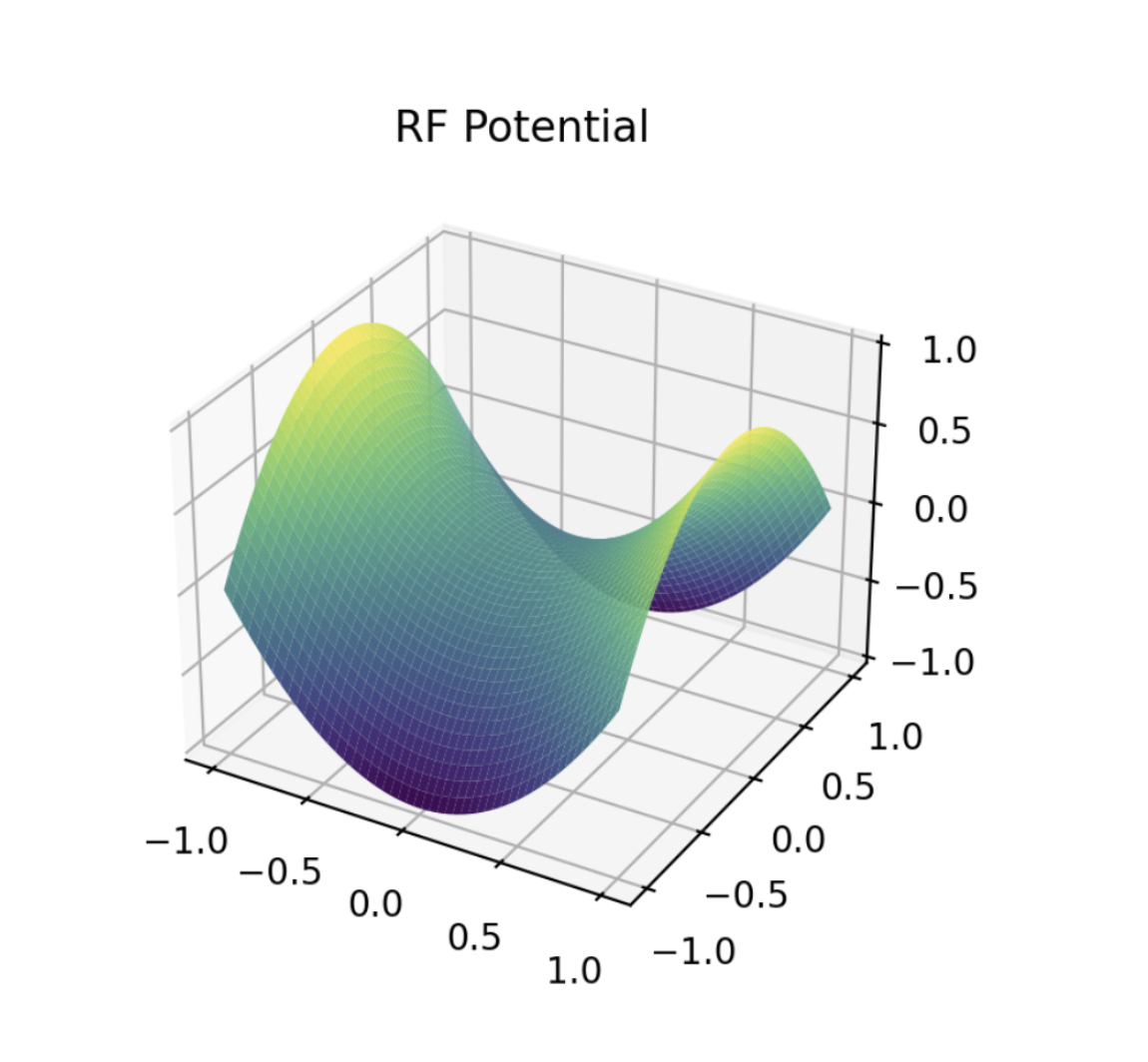}
    \caption{Visualization of the trapping potential.}
    \label{fig:sim_potential}
\end{figure}

\vspace{1em}In this equation, $U$ describes the DC voltage, $\tilde{U}$ the RF voltage, and $\Omega_{\text{rf}}$ the RF frequency. The first term represents the static confinement from the DC electrodes, while the second describes the oscillation generated by the RF drive. A simplified visualization of this alternating field is shown in Fig.~\ref{fig:sim_potential}.

This saddle-shaped potential flaps due to the oscillating nature of the RF field. This is the main force that allows the confinement of ions. Other aspects of ion trapping include the lasers used to load the ions and cool them down so that their kinetic motion is low enough to be trapped. Other than that, there is the vacuum system, whose role is to prepare and maintain an Ultra-High-Vacuum (UHV) environment, indispensable for ion trapping and the maintaining of coherence of the quantum states. Another subsystem to note is the imaging system. An imaging system often consists of a photomultiplier tube (PMT), a camera (cooled or CCD), and an objective lens. These are all constructed into an optical system and aligned appropriately to collect the fluorescence from atoms and ions within the vacuum chamber and image them through signals and images. Tab~\ref{tab:comp_sum} summarizes all of the core hardware components that a trapped-ion system consists of. Each of these modules plays a crucial role in ion trapping but is also the source of failure, as will be discussed in the following sections.

\begin{table}[htbp]
    \centering
    \caption{Core Hardware Components of Ion Trapping}
    \begin{tabular}{@{}p{1.2cm} p{7cm}@{}}
        \toprule
        \textbf{Module} & \textbf{Role} \\
        \midrule
        Vacuum & Achieves and maintains ultra-high vacuum (UHV, typically $<10^{-6}$ Pa) to minimize background gas collisions and maintain quantum coherence. \\
        \midrule
        Electronics &  Generates and controls the DC and RF voltages necessary for trapping ions with stability and minimum noise. \\
        \midrule
        Optics & Provide the required lasers for ablation (atom loading), photoionization, Doppler cooling, and internal state control, each at precisely tuned frequencies. \\
        \midrule
        Imaging & Collects and detects fluorescence emitted by the ion using objective lenses, cameras, and photomultiplier tubes (PMTs), often coupled with filters to isolate the signal from background noise. \\
        \bottomrule
    \end{tabular}
    \label{tab:comp_sum}
\end{table}

\section{Categorization of Failure Modes in Ion-Trapping}
\label{sec:fail_categ}
\begin{figure}[H]
    \centering
    \includegraphics[width=0.5\textwidth]{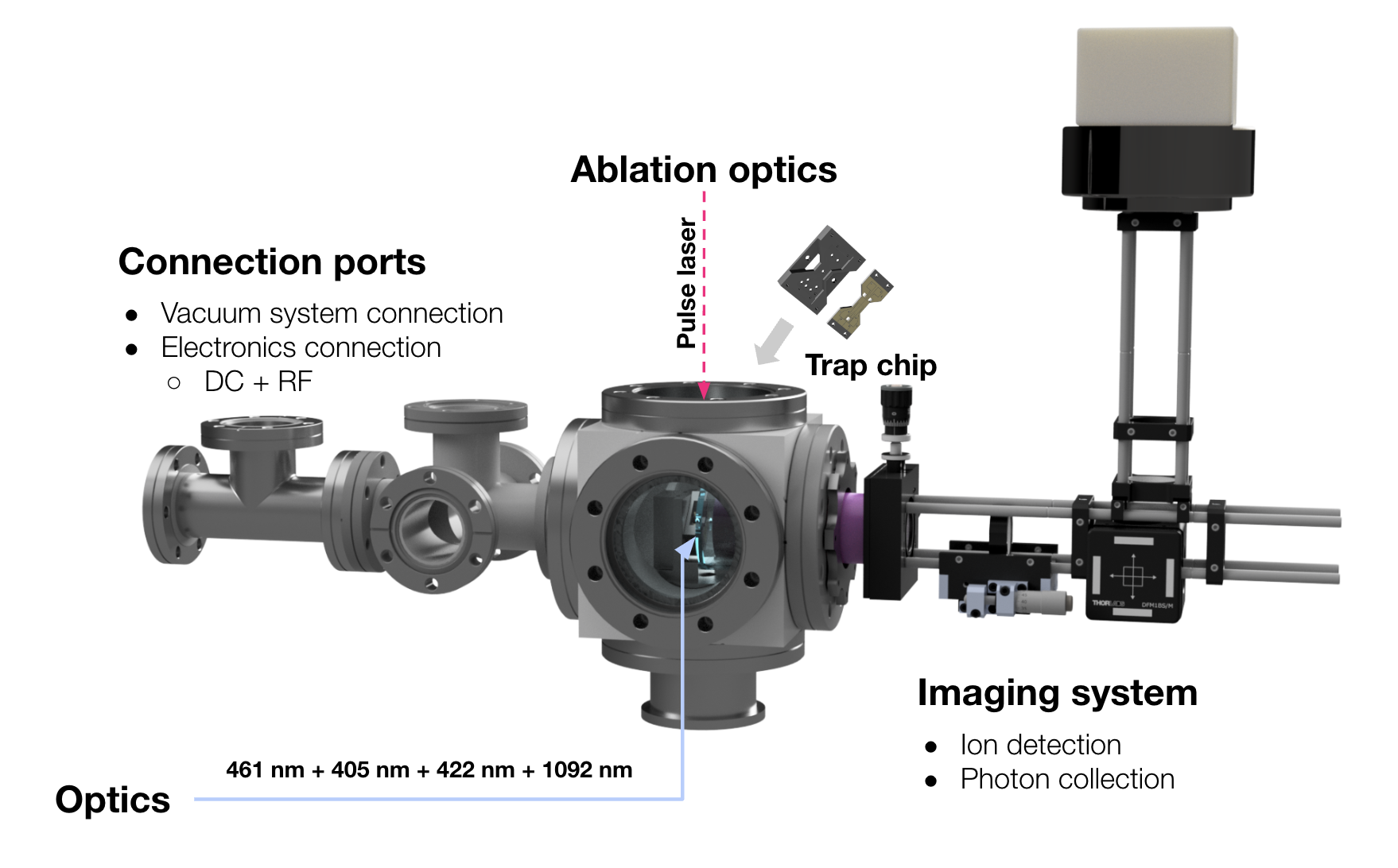}
    \caption{System overview of an ion-trapping system.}
    \label{fig:system_real}
\end{figure}

As discussed in Sec.~\ref{sec:basic_hardware}, multiple modules are involved in an ion-trap system. A collection of these subsystems is what allows for the trapping of ions and high-fidelity operations, which will be developed later in the construction of ion trap quantum computers. A simplified diagram of the ion-trapping system in our lab is shown in Fig~\ref{fig:system_real}.

A fluorescence signal detected from ions is what indicates whether ions are trapped or not. Once we find this signal through the PMT, we adjust the DC voltage configuration and RF voltages to maximize it. The next step would be to adjust the focal length of the objective lens (imaging system) to find the optimal focus of the trapped ions. However, even after going through all of the construction steps aimed towards trapping ions, it is not uncommon that you still can't find the trapping signal. When you are lost and do not know why you are not seeing the single, is when the most difficult part of ion-trapping begins —\textit{troubleshooting}. Through our experience in the lab, we categorized the main faults that could be intervening with your success.

\vspace{1em}\begin{enumerate}
\setlength\itemsep{1em}
    \item \textbf{Insufficient Ultra-High Vacuum (UHV):} 
    Achieving and maintaining a UHV environment is necessary for ion trapping. Failures here often come from leaks, inefficient bake-out, outgassing, or the malfunction of components within the vacuum system (e.g, Turbo pump, ion gauge).

    \item \textbf{Inadequate Trapping Potential:} 
    Trapping ions for an extended period highly depends on the stability of the trapping potential. The trapping potential is generated through the DC and RF electrodes on the ion trap. Issues can arise from the DC and RF sources malfunctioning or incorrect wiring of the electrodes, inefficient tuning of the RF frequency and power, or insufficient configuration of the DC voltages. Unwanted charge-ups of electrodes introduced from outgassing or electrical heating could cause unwanted fluctuations.

    \item \textbf{Ion Loading Failures:} 
    To trap ions, we need to have ions in the system. The process of loading the ions consisted of loading atoms first, then ionizing them. For ion trapping, there are a few methods for loading the atoms into the chamber, such as using ovens or a method called laser ablation. To ionize the loaded atoms, photoionization is commonly used.
    Ineffective loading of the atoms or the ionization lasers not being properly aligned could be the cause of the user not succeeding in ion trapping.

    \item \textbf{Ion Cooling Issues:} 
    After successfully loading ions, the next step is to slow them down to effectively trap them. The process of slowing them down utilizes a technique called laser cooling~\cite{Eschner:03}. The problems that could occur here are caused by misalignment of the laser and frequency detuning.

    \item \textbf{Imaging System Problems:} 
    Reliable ion detection depends on the imaging system. Issues include misalignment of the objective lens, such as incorrect focusing, or low signal-to-noise ratio due to environmental/stray light and camera or photomultiplier tube (PMT) malfunctions. The PMT is used to detect fluorescence from the atoms and ions in multiple stages of the construction and operation of trapped-ion systems.

    \item \textbf{Calibration Errors:} 
    Fine-tuning all of the components is necessary to trap ions. This includes RF signal tuning, DC structure, optical alignment, frequency tuning, and alignment of the imaging system. Once the engineer has a trapping signal, they may fine tune the DC voltages and RF source to maximize the signal.
\end{enumerate}

\vspace{1em}Note here that the main failure modes change system by system and the technology used. For example, for atom loading in our system, we use a method called laser ablation~\cite{osada2021loading}, while other groups may use different ways~\cite{Schioppo_2012}.

\section{Troubleshooting Framework}
\label{sec:trashoo}
The troubleshooting framework developed in this work allows users with limited expertise in ion trapping to systematically address potential system failures through a hierarchical, decision-tree approach. The user begins at the main tree Fig.~\ref{fig:main-tree}, which evaluates the entire system status from a higher level and guides toward module-specific subtrees (e.g., vacuum, optics, electronics) based on observed results at each step. Each node in the decision tree represents either an action (e.g., alignment) or a decision point that guides the user to different paths to support downstream analysis and evaluation. 
Unless stated otherwise, the term \textbf{signal} throughout this framework/paper refers to the fluorescence signal detected by a photomultiplier tube (PMT), which serves as the primary indicator of identifying success in specific steps and overall ion trapping. The main failure modes and the associated costs in each module are categorized in Section~\ref{sec:impact_eval}, Tab.~\ref{tab:unified_fmea_grouped}.

\subsection{Main Tree}
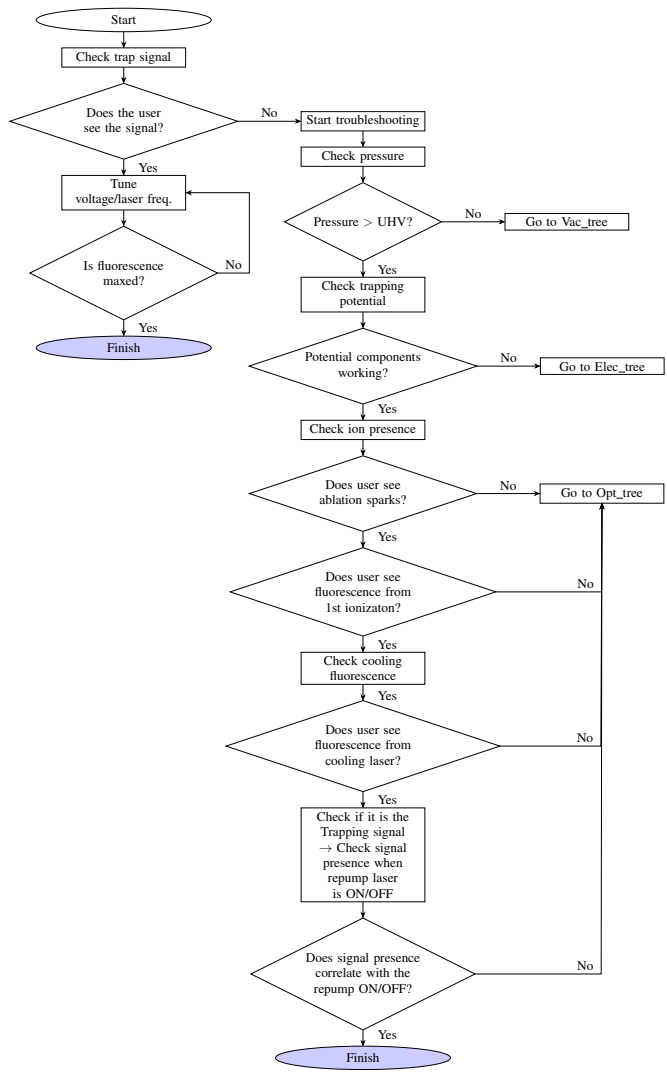
\begin{figure}[htbp]
    \centering
    \resizebox{0.48\textwidth}{!}{%
    \begin{tikzpicture}[node distance=0.5cm, font=\large, every node/.style={text width=3.7cm, align=center}, every path/.style={draw, -{Stealth}}]

        \node[ellipse, draw] (start) {Start};
        \node[rectangle, draw, below=of start] (checkSignal) {Check trap signal};
        \node[diamond, draw, below=of checkSignal, aspect=3] (signal) {Does the user see the signal?};
        \node[rectangle, draw, below=of signal] (tune) {Tune voltage/laser freq.};
        \node[diamond, draw, below=of tune, aspect=2] (fluorescence) {Is fluorescence maxed?};
        \node[ellipse, draw, below=of fluorescence, fill=blue!20] (finish) {Finish};
        \node[rectangle, draw, right=2cm of signal] (troubleshoot) {Start troubleshooting};

        \node[rectangle, draw, below=of troubleshoot] (pressure) {Check pressure};
        \node[diamond, draw, below=of pressure, aspect=2] (uhv) {Pressure $>$ UHV?};

        \node[rectangle, draw, below=of uhv] (potential) {Check trapping potential};
        \node[diamond, draw, below=of potential, aspect=3] (components) {Potential components working?};

        \node[rectangle, draw, below=of components] (ionPresence) {Check ion presence};
        \node[rectangle, draw, right=2cm of components] (elecTree) {Go to Elec\_tree};
        \node[rectangle, draw, right=2cm of uhv] (vacTree) {Go to Vac\_tree};
        \node[diamond, draw, below=of ionPresence, aspect=3] (ablationSpark) {Does user see ablation sparks?};
        \node[rectangle, draw, right=2cm of ablationSpark] (optTree) {Go to Opt\_tree};
        \node[diamond, draw, below=of ablationSpark, aspect=3] (1st_ionka) {Does user see fluorescence from 1st ionizaton?};
        \node[rectangle, draw, below=of 1st_ionka] (check_cool) {Check cooling fluorescence};
        \node[diamond, draw, below=of check_cool, aspect=3] (cool_fluor) {Does user see fluorescence from cooling laser?};

        \node[rectangle, draw, below=of cool_fluor] (checkTrap) {Check if it is the Trapping signal\\ $\rightarrow$ Check signal presence when repump laser is ON/OFF};
        \node[diamond, draw, below=of checkTrap, aspect=2] (check_repump) {Does signal presence correlate with the repump ON/OFF?};
        \node[ellipse, draw, below=of check_repump, fill=blue!20] (finish2) {Finish};

        \draw (start) -- (checkSignal);
        \draw (checkSignal) -- (signal);
        \draw (signal) -- node[right, xshift=-1.2cm]{Yes} (tune);
        \draw (signal) -- node[above]{No} (troubleshoot);
        \draw (tune) -- (fluorescence);
        \draw (fluorescence) -- node[right, xshift=-1.2cm]{Yes} (finish);
        \draw (fluorescence.east) -- ++(1,0) node[above, xshift=-0.5cm]{No} |- (tune.east);

        \draw (troubleshoot) -- (pressure);
        \draw (pressure) -- (uhv);
        \draw (uhv) -- node[right, xshift=-1.2cm]{Yes} (potential);
        \draw (uhv) -- node[above]{No} (vacTree);
        \draw (potential) -- (components);
        \draw (components) -- node[right, xshift=-1.2cm, yshift=0.1cm]{Yes} (ionPresence);
        \draw (components) -- node[above]{No} (elecTree);
        \draw (ionPresence) -- (ablationSpark);

        \draw (ablationSpark) -- node[above]{No} (optTree);

        \draw (ablationSpark) -- node[right, xshift=-1.2cm, yshift=0.1cm]{Yes} (1st_ionka);
        \draw (1st_ionka) -- node[right, xshift=-1.2cm]{Yes} (check_cool);
        \draw (check_cool) -- node[right, xshift=-1.2cm, yshift=-0.1cm]{Yes} (cool_fluor);
        \draw (cool_fluor) -- node[right, xshift=-1.2cm]{Yes} (checkTrap);
        \draw (checkTrap) -- (check_repump);
        \draw (check_repump) -- node[right, xshift=-1.2cm, yshift=0.1cm]{Yes} (finish2);
        \draw (1st_ionka.east) -- ++(3.35,0) node[above, xshift=-0.5cm]{No} -- (optTree.south);
        \draw (cool_fluor.east) -- ++(3.2,0) node[above, xshift=-0.5cm]{No} -- (optTree.south);
        \draw (check_repump.east) -- ++(4,0) node[above, xshift=-0.5cm]{No} -- (optTree.south);
    \end{tikzpicture}%
    }
    \caption{Main Tree: The starting point of troubleshooting an ion trap system.}
    \label{fig:main-tree}
\end{figure}

The main tree Fig.~\ref{fig:main-tree} is designed to survey each failure mode, introduced in Section~\ref{sec:fail_categ} to identify which area is causing trouble. Although the tree does not go into deep diagnosis, the user will most likely not go through this tree without being obstructed by one of the decision nodes, leading them to a sub-module/tree.

\subsection{Vacuum Tree}
There are different degrees of vacuum; however, for ion trapping, we want to operate in the UHV regime. To trap ions and use them as qubits for high-fidelity operations, we need to maintain a clean environment where few particles are present other than the ions being trapped. The degree of vacuum needed can differ by system, but it is recommended to be around $10^{-6}$ to $10^{-7}$ Pa.

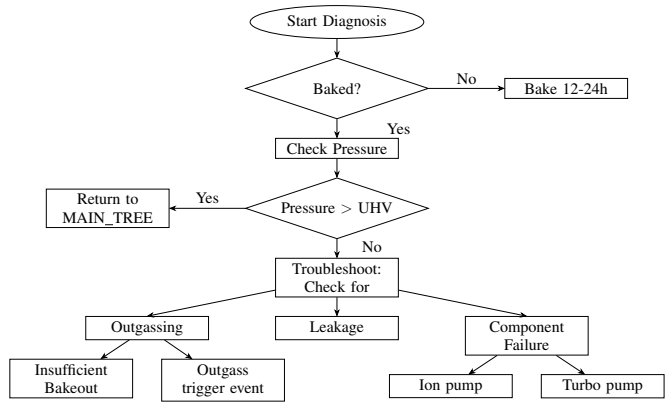
\begin{figure}[htbp]
    \centering
    \resizebox{0.48\textwidth}{!}{%
    \begin{tikzpicture}[node distance=0.5cm, font=\large, every node/.style={text width=3cm, align=center}, every path/.style={draw, -{Stealth}}]

        \node[ellipse, draw] (start) {Start Diagnosis};
        \node[diamond, draw, below=of start, aspect=3] (bake) {Baked?};
        \node[rectangle, draw, right=2cm of bake] (bake_action) {Bake 12-24h};
        \node[rectangle, draw, below=of bake] (check_pressure) {Check Pressure};
        \node[diamond, draw, below=of check_pressure, aspect=3] (uhv) {Pressure $>$ UHV};

        \node[rectangle, draw, below=of uhv] (troubleshoot) {Troubleshoot: Check for};
        \node[rectangle, draw, below= of troubleshoot] (leakage) {Leakage};
        \node[rectangle, draw, below=of troubleshoot, xshift=-5cm] (outgass) {Outgassing};
        \node[rectangle, draw, below=of troubleshoot, xshift=5cm] (component) {Component Failure};

        \node[rectangle, draw, below=of outgass, xshift=-2cm] (bad_bake) {Insufficient Bakeout};
        \node[rectangle, draw, below=of outgass, xshift=2cm] (gas_trigger) {Outgass trigger event};

        \node[rectangle, draw, below=of component, xshift=2cm] (turbo) {Turbo pump};
        \node[rectangle, draw, below=of component, xshift=-2cm] (ionpump) {Ion pump};
        \node[rectangle, draw, left=2cm of uhv] (return) {Return to MAIN\_TREE};

        \draw (start) -- (bake);
        \draw (bake) -- node[right]{Yes}(check_pressure);
        \draw (bake) -- node[above]{No}(bake_action);
        \draw (check_pressure) -- (uhv);
        \draw (uhv) -- node[above]{Yes}(return);
        \draw (uhv) -- node[right, xshift=-0.7cm, yshift=0.05cm]{No}(troubleshoot);
        \draw (troubleshoot) -- (leakage);
        \draw (troubleshoot) -- (outgass.north);
        \draw (outgass) -- (bad_bake);
        \draw (outgass) -- (gas_trigger);
        \draw (troubleshoot) -- (component.north);
        \draw (component) -- (ionpump);
        \draw (component) -- (turbo);
    \end{tikzpicture}%
    }
    \caption{Vacuum\_Tree: Troubleshooting flow for the vacuum system.}
    \label{fig:vac_tree}
\end{figure}

In addition to the main failure modes categorized after the \textit{Troubleshooting action node} in Fig.~\ref{fig:vac_tree}, there is still the risk of measurement fault. For instance, if the ion gauge —used to measure the pressure of the inside of a vacuum chamber— is malfunctioning, the user could misdiagnose the state of the vacuum environment. This will lead to the false assumption that the system has reached the UHV regime when, in fact, it has not, ultimately making ion trapping impossible. When troubleshooting trapped-ion systems, make sure to second-guess the baseline components that users usually trust.

Among vacuum module failure modes, outgassing is something that cannot be entirely avoided. Outgassing occurs when residual gases are released from materials inside the vacuum chamber.  A summary of the causes and their descriptions are given in Tab~\ref{tab:outgas_cause}.

\begin{table}[htbp]
    \centering
    \caption{Causes of Outgassing in Vacuum Systems}
    \begin{tabular}{@{} p{3cm} p{5cm} @{}}
        \toprule
        \textbf{Cause} & \textbf{Description} \\
        \midrule
        Ineffective Bake-Out & Inadequate baking leave residual gases adsorbed on surfaces. \\
        \midrule
        Trigger Events & Events such as laser ablation introduce gases in the chamber. \\
        \midrule
        Electrical Heating & Heat from electrical components releases trapped gases. \\ 
        \midrule
        Non-vacuum-compatible materials & Unsuitable materials inside the chamber continually release gas. \\
        \bottomrule
    \end{tabular}
    \label{tab:outgas_cause}
\end{table}

\subsection{Electronics Tree}
The electronic module responsible for preparing a stable trapping potential consists of two main areas: DC electronics and RF (radio frequency) electronics. 

\begin{figure}[htbp]
    \centering
    \resizebox{0.48\textwidth}{!}{%
    \begin{tikzpicture}[node distance=0.5cm, font=\large, every node/.style={text width=3cm, align=center}, every path/.style={draw, -{Stealth}}]

        \node[ellipse, draw] (start) {Start Diagnosis};
        \node[diamond, draw, below=of start, aspect=3] (source_stat) {Is source ON?};
        \node[rectangle, draw, below=of source_stat] (troublebranch) {Troubleshoot Path};
        \node[rectangle, draw, below=of troublebranch, xshift=-5.2cm] (dc) {DC};
        \node[rectangle, draw, below=of troublebranch, xshift=5.2cm] (rf) {RF};
        \node[rectangle, draw, right=5cm of source_stat] (turn_on) {Turn source ON and proceed};

        \node[rectangle, draw, below=of rf] (helical) {Resonator reflectance};
        \node[rectangle, draw, below=of helical] (check_coupling) {Check coupling\\ $\rightarrow$VNA $+$ resonator};
        \node[diamond, draw, below=of check_coupling, aspect=2] (reflect0) {Is RF critically coupled/\\ Is reflectance 0?};
        \node[rectangle, draw, right=of reflect0] (helical_couple) {Adjust helical resonator to achieve critical coupling.};
        \node[rectangle, draw, below=of reflect0] (rf_config) {Configure to optimal parameters};
        \node[rectangle, draw, below=of rf_config, xshift=-3cm] (rf_freq) {Frequency};  
        \node[rectangle, draw, below=of rf_config, xshift=3cm] (rf_pow) {Power};

        \node[rectangle, draw, below=of dc] (dc_connect) {Connection};
        \node[rectangle, draw, below=of dc_connect] (check_net) {Check network connection \\ $\rightarrow$ CP connected to DAC};
        \node[diamond, draw, below=of check_net, aspect=3] (net_stable) {Is connection stable?};
        \node[rectangle, draw, right=of net_stable] (dc_fix0) {Further diagnosis on connection.};
        \node[rectangle, draw, below=of net_stable] (check_dc_out) {Check DC voltage at output (filter)};
        \node[diamond, draw, below=of check_dc_out, aspect=3] (proper_volt) {Is voltage being sent correctly?};
        \node[rectangle, draw, below=of proper_volt] (check_dc_config) {Check the DC voltage configuration sent to chip};
        \node[diamond, draw, below=of check_dc_config, aspect=3] (dc_config_neat) {Is the DC configured in a balanced matter?};
        \node[rectangle, draw, below=of dc_config_neat] (dc_return) {Go back to MAIN\_TREE};
        \node[rectangle, draw, right=of dc_config_neat] (mk_dc_balance) {Configure DC voltages in a balanced way};

        \draw (start) -- (source_stat);
        \draw (source_stat) -- node[right, xshift=-1.2cm]{Yes} (troublebranch);
        \draw (troublebranch) -- (dc.north);
        \draw (troublebranch) -- (rf.north);
        \draw (source_stat) -- node[above]{No} (turn_on);
        \draw (dc) -- (dc_connect);
        \draw (dc_connect) -- (check_net);
        \draw (check_net) -- (net_stable);
        \draw (net_stable) -- node[right, xshift=-1.2cm]{Yes} (check_dc_out);
        \draw (net_stable) -- node[above, xshift=-0.1cm]{No} (dc_fix0);
        \draw (check_dc_out) -- (proper_volt);
        \draw (proper_volt) -- node[right, xshift=-1.2cm]{Yes} (check_dc_config);
        \draw (check_dc_config) -- (dc_config_neat); 
        \draw (dc_config_neat) -- node[above, xshift=-0.1cm]{No} (mk_dc_balance);
        \draw (dc_config_neat) -- node[right, xshift=-1.2cm]{Yes} (dc_return);
        
        \draw (rf) -- (helical);
        \draw (helical) -- (check_coupling);
        \draw (check_coupling) -- (reflect0);
        \draw (reflect0) -- node[above, xshift=-0.2cm]{No} (helical_couple);
        \draw (reflect0) -- node[right, xshift=-1.2cm, yshift=0.1cm]{Yes} (rf_config);
        \draw (rf_config) -- (rf_freq);
        \draw (rf_config) -- (rf_pow);
        
    \end{tikzpicture}%
    }
    \caption{Electronics Tree: Troubleshooting flow for the electronic system, comprehensive to DC and RF components.}
    \label{fig:elc_tree}
\end{figure}
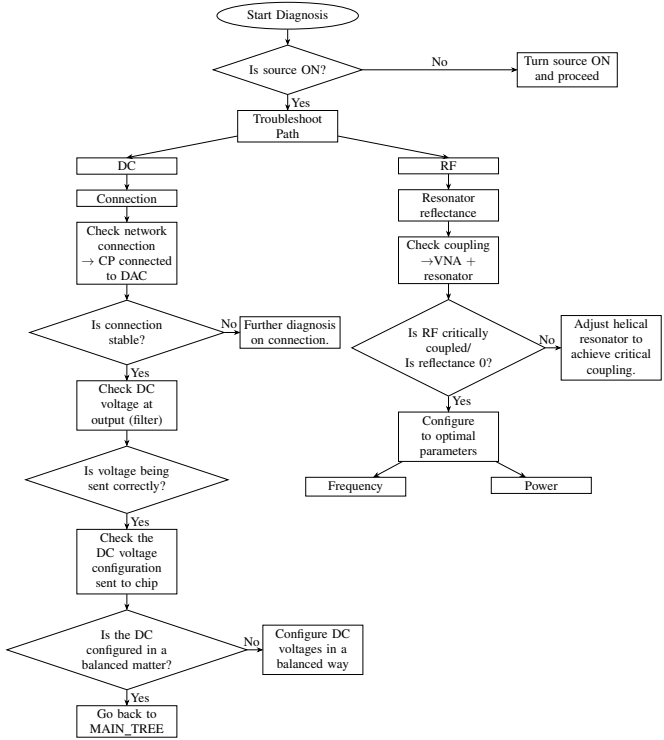

A balanced DC configuration refers to setting the DC voltages on each of the trap-chip electrodes in a geometrically balanced way, rather than using uneven or arbitrary values. Even if the optimal DC voltage configuration is not yet set for the user's system or specific trap-chip, a even voltage is a safe and stable starting point.

\subsection{Optics Tree}
The optics module includes all lasers involved in the loading and cooling of ions. These include the ablation laser, ionization laser, and the cooling and repump lasers. The optics subsystem can be subdivided into these three sections. When the user is at the stage of system construction that includes quantum gates, you will need to add the lasers for quantum operation and sideband cooling to cool the ions to their emotional ground state.

\begin{figure}[htbp]
    \centering
    \resizebox{0.48\textwidth}{!}{%
    \begin{tikzpicture}[node distance=0.5cm, font=\large, every node/.style={text width=3.5cm, align=center}, every path/.style={draw, -{Stealth}}]

        \node[ellipse, draw] (start) {Start Diagnosis};
        \node[rectangle, draw, below=of start] (check_cam) {Check camera};
        \node[diamond, draw, below=of check_cam, aspect=2] (spark) {Do you see ablation sparks?};
        \node[rectangle, draw, below=of spark] (ionization) {Ionization};
        \node[rectangle, draw, right=of spark, xshift=3cm] (atom_load) {Ablation};

        \node[rectangle, draw, below=of ionization] (check_pmt) {Check fluorecence\\ $\rightarrow$ PMT w/ bandpass filter};
        \node[diamond, draw, below=of check_pmt, aspect=2] (1st_fluor) {Do you see signal from 1st ionization laser?};
        \node[rectangle, draw, right=of 1st_fluor, xshift=2cm] (troubleshoot) {Troubleshoot};
        \node[rectangle, draw, below=of troubleshoot, xshift=-2cm] (align) {Alignment};
         \node[rectangle, draw, below=of troubleshoot, xshift=2cm] (laser_freq) {Laser frequency};
        \node[rectangle, draw, below=of 1st_fluor] (check_lap) {Check laser overlap};
        \node[diamond, draw, below=of check_lap, aspect=2] (laser_lap) {Are 1st and 2nd ionization lasers overlapping?};
        \node[rectangle, draw, below=of laser_lap] (2nd_freq) {Check frequency of 2nd ionization laser};
        \node[rectangle, draw, right=of laser_lap, xshift=2.3cm] (align2) {Align laser paths};
        \node[diamond, draw, below=of 2nd_freq, aspect=2] (2nd_freq_match) {Does frequency match transition of 2nd ionization?};
        \node[rectangle, draw, right=of 2nd_freq_match, xshift=2.3cm] (adjust_freq) {Adjust laser frequency};

        \node[rectangle, draw, below=of 2nd_freq_match] (cool) {Cooling};
        \node[rectangle, draw, below=of cool] (precon) {Precondition:Ions are loaded};
        \node[rectangle, draw, below=of precon, xshift=-3cm] (align3) {Alignment};
        \node[rectangle, draw, below=of precon, xshift=3cm] (cool_freq) {Laser frequency};
        \node[rectangle, draw, below=of align3] (check_cool_lap) {Check laser overlap};
        \node[diamond, draw, below=of check_cool_lap, aspect=2] (cool_lapping) {Is the cooling laser overlapping with the ionization lasers?};
        \node[rectangle, draw, below=of cool_lapping] (next_katei) {Proceed to next failure mode};
        \node[rectangle, draw, right=of cool_lapping] (align4) {Align cooling laser and repump laser to overlap w/ ionization laser};

        \node[rectangle, draw, below=of atom_load, xshift=-3cm] (alignment) {Alignment};
        \node[rectangle, draw, below=of atom_load, xshift=3cm] (laser_calib) {Laser calibration};

        \draw (start) -- (check_cam);
        \draw (check_cam) -- (spark);
        \draw (spark) -- node[above, xshift=-1.2cm]{Yes} (ionization);
        \draw (spark) -- node[above]{No} (atom_load);
        \draw (ionization) -- (check_pmt);
        \draw (check_pmt) -- (1st_fluor);
        \draw (1st_fluor) -- node[right, xshift=-1.2cm]{Yes} (check_lap);
        \draw (1st_fluor) -- node[above]{No} (troubleshoot);
        \draw (check_lap) -- (laser_lap);
        \draw (laser_lap) -- node[above]{No} (align2);
        \draw (align2.north) -- ++(0,0) |- (check_lap.east);
        \draw (laser_lap) -- node[right, xshift=-1.2cm]{Yes} (2nd_freq);
        \draw (2nd_freq) -- (2nd_freq_match);
        \draw (2nd_freq_match) -- node[above]{No} (adjust_freq);
        \draw (2nd_freq_match) -- node[right, xshift=-1.2cm]{Yes} (cool);
        \draw (troubleshoot) -- (laser_freq);
        \draw (troubleshoot) -- (align);
        \draw (cool) -- (precon);
        \draw (precon) -- (align3);
        \draw (precon) -- (cool_freq);
        \draw (align3) -- (check_cool_lap);
        \draw (check_cool_lap) -- (cool_lapping);
        \draw (cool_lapping) -- node[right, xshift=-1.2cm, yshift=0.1cm]{Yes} (next_katei);
        \draw (cool_lapping) -- node[above, xshift=-0.1cm]{No} (align4);
        
        \draw (atom_load) -- (alignment);
        \draw (atom_load) -- (laser_calib);
    \end{tikzpicture}%
    }
    \caption{Optics Tree: Simplified due to the size of the full optics module.}
    \label{fig:simp_opt_tree}
\end{figure}
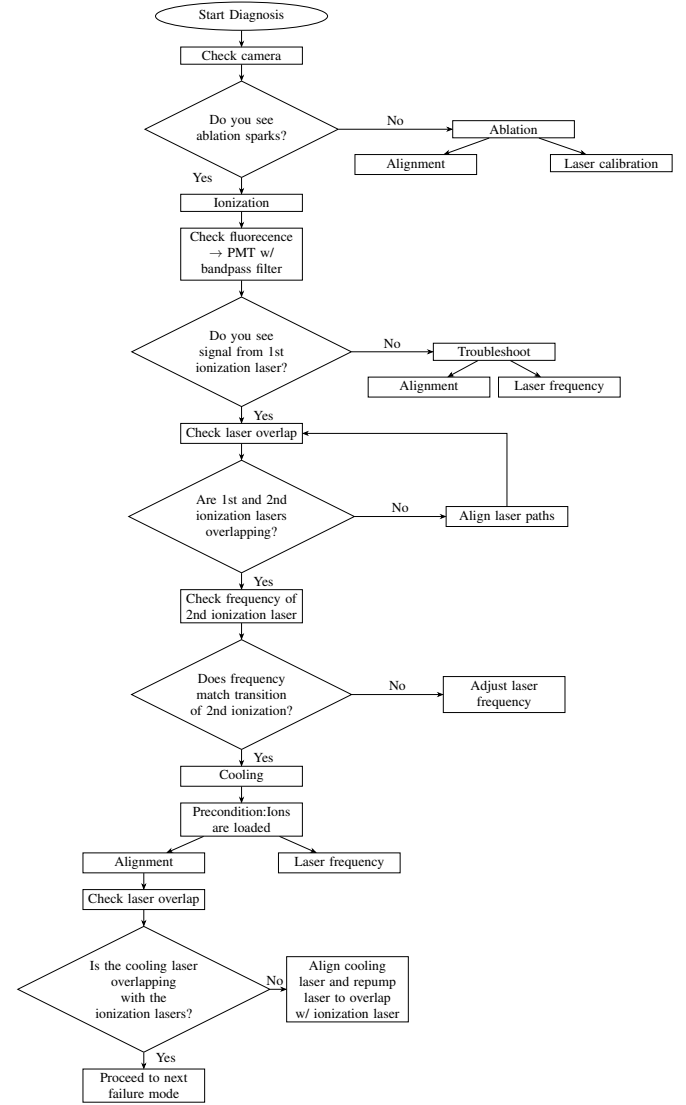

The areas of fault when it comes to optics are mainly the alignment and frequencies of the lasers. The lasers are used to drive a certain transition within the specific ion species that the practitioner wants to trap. To drive the transition effectively, we need to lock the laser frequency at a specific point that has a very narrow window.

The optics module is largely divided into two tracks, Ablation and Ionization/Cooling.
If ablation is not working, Ionization and cooling are off-limits, however, it is relatively easy to evaluate whether ablation is working or not. Due to the size of this full module, we will split the module tree into two sections for visibility purposes. For the ablation laser, in our case, the wavelength of the laser was set, so the variables that we could control were:

\vspace{1em}\begin{itemize}
\setlength\itemsep{1em}
    \item \textbf{Alignment:} Misalignment can result in insufficient material being ablated, leading to a lack of atoms in the trapping region. 
    
    \item \textbf{Laser Calibration:} Low energy or improper pulse settings may fail to release enough atoms, while excessive energy can damage the target or lead to inefficient ionization. 
\end{itemize}

\begin{figure}[htbp]
    \centering
    \resizebox{0.48\textwidth}{!}{%
    \begin{tikzpicture}[node distance=0.5cm, font=\large, every node/.style={text width=3cm, align=center}, every path/.style={draw, -{Stealth}}]
        \node[rectangle, draw] (atom_load) {Ablation};
        \node[rectangle, draw, below=of atom_load] (troublebranch) {Troubleshoot paths};
        \node[rectangle, draw, below=of troublebranch, xshift=-5cm] (alignment) {Alignment};
        \node[rectangle, draw, below=of alignment] (check_path) {Check path\\ $\rightarrow$ w/ visible light};
        \node[diamond, draw, below=of check_path, aspect=2] (abl_path_good) {Is laser path hitting target?};
        \node[rectangle, draw, right=of abl_path_good] (adjust_abl) {Adjust ablation path};
        \node[diamond, draw, below=of abl_path_good, aspect=2] (abl_path_good2) {Is laser path in range of trapping reagion?};

        \node[rectangle, draw, below=of troublebranch, xshift=5cm] (laser_calib) {Laser calibration};
        \node[rectangle, draw, below=of laser_calib, xshift=-2cm] (laser_pow) {Power};
        \node[rectangle, draw, below=of laser_calib, xshift=2cm] (pulse_freq) {Pulse freqency};
        \node[diamond, draw, below=of laser_calib, yshift=-1cm, aspect=2] (abl_pulse_good) {Is setting optimal?};
        \node[rectangle, draw, right=of abl_pulse_good] (pulse_calib) {Calibrate ablation laser source};

         \node[rectangle, draw, below=of troublebranch, yshift=-10cm] (check_cam2) {Check camera for ablation spark};
         \node[diamond, draw, below=of check_cam2, aspect=2] (look4sparks) {Are you seeing ablation sparks?};
         \node[rectangle, draw, right=of look4sparks] (other_branch) {Diagnose other branch};
         \node[ellipse, draw, below=of look4sparks, fill=blue!20] (goback) {Return to Optics Module};

        \draw (atom_load) -- (troublebranch);
        \draw (troublebranch) -- (alignment);
        \draw (alignment) -- (check_path);
        \draw (check_path) -- (abl_path_good);
        \draw (abl_path_good) -- node[right, xshift=-1.2cm, yshift=0.1cm]{Yes} (abl_path_good2);
        \draw (abl_path_good) -- node[above, xshift=-0.1cm]{No} (adjust_abl);
        \draw (troublebranch) -- (laser_calib);
        \draw (laser_calib) -- (laser_pow);
        \draw (laser_calib) -- (pulse_freq);
        \draw (laser_pow) -- (abl_pulse_good.north);
        \draw (pulse_freq) -- (abl_pulse_good.north);
        \draw (abl_pulse_good.south) -- ++(0,0) node[above, xshift=-2cm, yshift=-5.5cm]{Yes} |- (check_cam2);
        \draw (abl_pulse_good) -- node[above, xshift=-0.1cm]{No} (pulse_calib);
        \draw (abl_path_good2.south) -- ++(0,0) node[above, xshift=2cm, yshift=-1cm]{Yes} |- (check_cam2);
        \draw (check_cam2) -- (look4sparks);
        \draw (look4sparks) -- node[right, xshift=-1.2cm, yshift=0.1cm]{Yes} (goback);
        \draw (look4sparks) -- node[above, xshift=-0.1cm]{No} (other_branch);
    \end{tikzpicture}%
    }
    \caption{Ablation track of the Optics Module.}
    \label{fig:abl_opt_tree}
\end{figure}
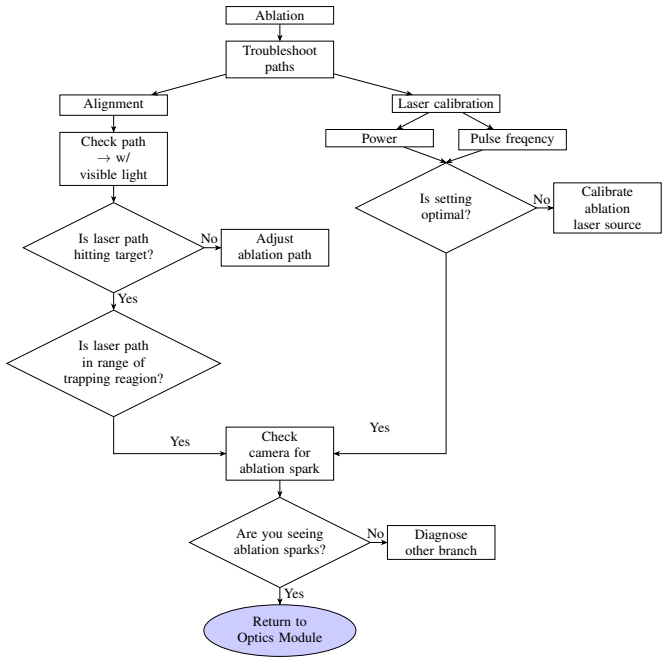

\begin{table}[htbp]
    \centering
    \caption{Diagnosis and Correction methods for laser ablation}
    \begin{tabular}{@{} p{3cm} p{5cm}@{}}
        \toprule
        \textbf{Method} & \textbf{Description} \\
        \midrule
        Check the laser path with visible light & Use a visible light alignment beam to ensure the laser path overlaps with the target ablation point. \\
        \midrule
        Check laser power and pulse frequency & Check laser controller to verify the power and pulse frequency. \\
        Realign the laser path & Use a visible light alignment beam to specify the target path. Adjust mirrors and lenses to ensure the ablation laser overlaps with the target. \\
        \midrule
        Optimize Power and Frequency & Incrementally adjust the laser power and pulse frequency while monitoring ablation sparks with a camera. \\
        \bottomrule
    \end{tabular}
    \label{tab:abl_diagnosis}
\end{table}

Choosing the optimal setting for the ablation laser depends on the laser in use in the engineer's setup. If it is the first time that you are working with that setup, the engineer will have to sweep parameters to find the best-performing poing. This node is just to notify the user that the status of ion loading could depend on the setting/calibration of the laser in use.

\subsection{Imaging Tree}
The imaging system, often already diagnosed at the beginning of the construction, could be the potential issue on why we are not detecting ions. The areas of errors in the imaging system are categorized in Tab~\ref{tab:unified_fmea_grouped}.

\begin{figure}[htbp]
    \centering
    \resizebox{0.48\textwidth}{!}{%
    \begin{tikzpicture}[node distance=0.5cm, font=\large, every node/.style={text width=3.5cm, align=center}, every path/.style={draw, -{Stealth}}]

        \node[ellipse, draw] (start) {Start Diagnosis};
        \node[rectangle, draw, below=of start] (check_sn) {Check SN ratio w/ PMT};
        \node[diamond, draw, below=of check_sn, aspect=3] (sn_min) {Is SN ratio $>$ 30};

        \node[rectangle, draw, below=of sn_min] (troubleshoot) {Troubleshoot: Check for};
        \node[rectangle, draw,  below=of troubleshoot, xshift=-4cm] (alignment) {Alignment};
        \node[rectangle, draw, below=of troubleshoot, xshift=4cm] (comp_fail) {Component Failure};

        \node[rectangle, draw, below= of alignment, xshift=2cm] (internal) {Internal alignment};
        \node[rectangle, draw, below=of alignment, xshift=-2cm] (external) {External alignment};

        \node[rectangle, draw, below= of comp_fail, xshift=2cm] (pmt) {PMT};
        \node[rectangle, draw, below=of comp_fail, xshift=-2cm] (camera) {Camera};

        \node[rectangle, draw, right=2cm of sn_min] (coverup) {Cover with light block material};

        \draw (start) -- (check_sn);
        \draw (check_sn) -- (sn_min);
        \draw (coverup.north) -- ++(1,0) node[above, xshift=-0.5cm]{No} |- (check_sn.east);
        \draw (sn_min) -- node[right, xshift=-1cm, yshift=0.1cm]{Yes}(troubleshoot);
        \draw (sn_min) -- node[above]{No}(coverup);
        \draw (troubleshoot) -- (alignment);
        \draw (troubleshoot) -- (comp_fail);
        \draw (alignment) -- (internal);
        \draw (alignment) -- (external);
        \draw (comp_fail) -- (pmt);
        \draw (comp_fail) -- (camera);
    \end{tikzpicture}%
    }
    \caption{Imaging module of the Troubleshooting Framework}
    \label{fig:img_tree}
\end{figure}
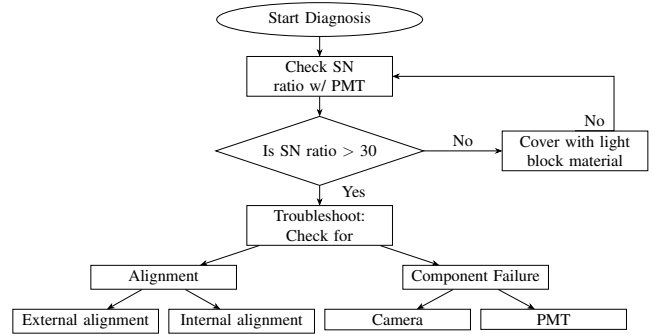

Here, internal alignment refers to the alignment of the imaging module itself. If this is not optimal, issues could occur, such as the detection area of the PMT and camera could be different.. External alignment, on the other hand, points to the alignment of the imaging module with the chamber/trap chip.

\section{Qualitative Impact Analysis of Ion Trap Failures}
\label{sec:impact_eval}

\begin{table*}[t]
    \centering
    \caption{Qualitative Impact Assessment of Representative Failure Modes.}
    \begin{tabular}{@{}p{1.5cm} p{4cm} p{3cm} p{3cm} p{3cm}@{}}
    \toprule
    \textbf{Area} & \textbf{Main Failure Modes} & \textbf{Operational Impact} & \textbf{Time Cost} & \textbf{Disturbance Risk} \\
    \midrule
    \multirow{3}{*}{Vacuum} 
        & Outgassing (bake-out failure) & High & High & Low \\
        & Leak (gasket/valve) & High & High & High \\
        & Component failure & High & High & Low \\
    \midrule
    \multirow{3}{*}{Electronics} 
        & RF detuning (resonator reflectance) & High & High & High \\
        & DC noise / voltage drift & Medium & Medium & Medium \\
        & Broken cable / poor contact & Medium & High & High \\
    \midrule
    \multirow{3}{*}{Optics} 
        & Ionization laser misalignment & Medium & Medium & Medium \\
        & Cooling beam misaligned & Medium & Medium & Low \\
        & Laser frequency drift & Low & Medium & Low \\
    \midrule
    \multirow{2}{*}{Imaging} 
        & Camera/PMT misalignment & Medium & Medium & Medium \\
        & Light leak / poor shielding & Low & Low & Low \\
    \bottomrule
\end{tabular}
\label{tab:unified_fmea_grouped}
\end{table*}

When designing reliable systems, we need to take into account the potential problems that can happen and categorize them by their cost so that we know what we are dealing with. To do so, we adopt the FMEA (Failure Mode and Effects Analysis) methodology~\cite{NASA_1966} to effectively categorize and evaluate common failure modes in the construction of an ion-trap system. While the cited document performs FMEACA (Failure Mode, Effects and Critical Analysis), which includes quantitative analysis based on statistical data, such analysis is not yet feasible with the setup that we have in our lab. For this paper, we keep it at a qualitative analysis that lays out the foundation for quantitative evaluation in the future.

\subsection{Evaluation Dimensions}
\label{sec:cost}
Before we categorize the failure modes and their associated costs, we need to define the dimensions of evaluation. Each failure mode is assessed along three axes:

\vspace{1em}\begin{enumerate}
\setlength\itemsep{1em}
    \item \textbf{Operational Impact:} How problematic the error is on the operation of the quantum node.
    \item \textbf{Time Cost:} Estimated human work time and experimental delay.
    \item \textbf{Disturbance Risk:} How much the fixing of the failure could introduce new trouble.
\end{enumerate}

\vspace{1em}Analyzing the cost of the main failure modes of each sub-module will allow users and engineers to prioritize where to build diagnostics, invest in automation, or redesign components.

\begin{table}[htbp]
    \centering
    \caption{Effect of Fault Severity Level}
    \begin{tabular}{@{} p{1.5cm} p{2cm} p{2cm} p{2cm}@{}}
        \toprule
        \textbf{Dimension} & \textbf{Low} & \textbf{Medium} & \textbf{High} \\
        \midrule
        Time Cost & Hours & Days & Weeks \\
        \midrule
        Operational Impact & Motion is introduced & Trapped for few minutes & Ion loss\\
        \midrule
        Misalignment Risk & Unlikely to happen & Could happen & Mostlikely will happen \\
        \bottomrule
    \end{tabular}
    \label{tab:effect_fault}
\end{table}

Table~\ref{tab:effect_fault} defines the effects of each failure mode's severity in three degrees.

\vspace{1em}\begin{table}[htbp]
    \centering
    \caption{Definition of Fault Severity Level and Interventions}
    \begin{tabular}{@{}p{1cm} p{3cm} p{4cm}@{}}
        \toprule
        \textbf{Severity} & \textbf{Definition} & \textbf{Typical Intervention} \\
        \midrule
        Low & Minor performance loss or cosmetic fault & Simple recalibration or adjustment \\
        \midrule
        Medium & Degrades fidelity or reliability of trapping & Re-alignment or moderate component replacement \\
        \midrule
        High & Prevents trapping or damages hardware & Extensive intervention (e.g. disassembly, replacing hardware) \\
        \bottomrule
    \end{tabular}
    \label{tab:severity_levels}
\end{table}

With all of the definitions out of the way, we can review the failure types per subsystem, which are categorized in Tab~\ref{tab:unified_fmea_grouped} with their qualitative assets cost as defined in Sec~\ref{sec:cost}.

\subsubsection{Vacuum module Failure Modes}
Leakage and component failure are categorized as high-cost failure modes, primarily because they often require hardware replacement. This introduces delays due to shipping times and the need to reopen the vacuum chamber. If ions are already trapped, such failures typically result in ion loss, making them critical for any ongoing quantum operations. However, the actual severity depends on the source of the issue: a cracked chamber would be catastrophic, while a worn gasket or imperfect seal might be resolved more easily, with only medium-level system impact.

\vspace{1em}\subsubsection{Electronic module Failure Modes}
RF detuning—though relatively rare—is classified as high in both operational impact and time cost. If unresolved, it can disrupt the trapping potential, ultimately leading to the loss of trapped-ions. The likelihood of this failure depends on the system's RF components (e.g., helical resonator) and wiring, which vary between setups.

\vspace{1em}\subsubsection{Optics module Failure Modes}
The primary failure modes in the optics module—frequency drift and misalignment—are generally correctable without affecting other subsystems. While realigning optics can be time-consuming, most of the effort lies in the initial build. Later adjustments typically require only fine-tuning, especially once the main optical paths have been established.

\vspace{1em}\subsubsection{Imaging module Failure Modes}
Imaging failures are less frequent but highly dependent on system design. A well-designed imaging system should allow mid-construction diagnostics and easy realignment. Since it is often placed close to the vacuum chamber, realigning this module can disturb nearby components, such as pre-aligned optical systems.

\subsection{System-Level Trends and Observations}
From the qualitative evaluation, we can observe the following trends.

\vspace{1em}\begin{itemize}
    \item \textbf{Vacuum and Electronics} failures are often the most critical, especially when involving hardware replacement or opening the vacuum chamber.
    
    \item \textbf{Optics} failures are frequent but typically less destructive; many can be corrected via alignment or laser tuning.

    \item \textbf{Imaging system} issues are hard to isolate due to indirect visibility; even small misalignments can be confusing.

    \item \textbf{Calibration errors} are common across all subsystems and are often low in operational impact but high in time cost due to trial-and-error processes.
\end{itemize}

\section{Discussion}
\label{sec:Discussion}
This work represents an early step toward engineering ion trap systems to take the experiments to systems and to make them into products that we can deploy and use at large scales. While trapped ions are a leading candidate for scalable quantum nodes, the lack of structured diagnostics and troubleshooting protocols has kept ion-trap systems within the laboratory, holding up efforts to build real-world systems/products that have reproducibility and maintenance in mind. Our framework addresses this gap by introducing a modular, subsystem-level troubleshooting methodology, inspired by principles from hardware reliability engineering such as FMEA. The proposed framework provides immediate value by making the troubleshooting process more transparent, reproducible, and accessible, particularly for teams where experimental physicists, engineers, and software developers must collaborate. 

There are limitations. The framework is currently tailored to a specific experimental setup and lacks quantitative fault data. As ion-trap systems are not yet standardized, system-specific variations make universal applicability challenging. However, this structured approach provides the necessary structure for broader adoption. As more groups use and refine this framework, community-driven extensions could generalize it across different ion species, trap designs, and control architectures. 

We envision extending this framework in the following directions such as:

\vspace{1em}\begin{itemize}
\setlength\itemsep{1em}
    \item \textbf{Design-for-debug:} Guide trap system design with failure mitigation and maintainability in mind.
    
     \item \textbf{Proactive error handling:} Integrate fault detection with automated recovery protocols (e.g., real-time laser tuning, alignment checks).
     
    \item \textbf{Open-source diagnostic knowledge base:} Enable labs to share structured failure cases and remediation paths, accelerating the development of error-aware architectures.
\end{itemize}

Ultimately, we hope this work contributes toward the development of robust, scalable, and debuggable quantum systems, laying the groundwork for hardware/software co-design and real-world deployment of quantum network nodes.

\bibliographystyle{IEEEtran.bst}
\bibliography{IEEEabrv, bibfile}

\begin{thebibliography}{10}
\providecommand{\url}[1]{#1}
\csname url@samestyle\endcsname
\providecommand{\newblock}{\relax}
\providecommand{\bibinfo}[2]{#2}
\providecommand{\BIBentrySTDinterwordspacing}{\spaceskip=0pt\relax}
\providecommand{\BIBentryALTinterwordstretchfactor}{4}
\providecommand{\BIBentryALTinterwordspacing}{\spaceskip=\fontdimen2\font plus
\BIBentryALTinterwordstretchfactor\fontdimen3\font minus \fontdimen4\font\relax}
\providecommand{\BIBforeignlanguage}[2]{{%
\expandafter\ifx\csname l@#1\endcsname\relax
\typeout{** WARNING: IEEEtran.bst: No hyphenation pattern has been}%
\typeout{** loaded for the language `#1'. Using the pattern for}%
\typeout{** the default language instead.}%
\else
\language=\csname l@#1\endcsname
\fi
#2}}
\providecommand{\BIBdecl}{\relax}
\BIBdecl

\bibitem{meter2006}
R.~D. Van{ }Meter{ }III, ``Architecture of a quantum multicomputer optimized for {Shor's} factoring algorithm,'' Ph.D. dissertation, Keio University, 2006, available as arXiv:quant-ph/0607065.

\bibitem{rfc9340}
\BIBentryALTinterwordspacing
W.~Kozlowski \emph{et~al.}, ``{Architectural Principles for a Quantum Internet},'' RFC 9340, Mar. 2023, \href{https://dx.doi.org/10.17487/RFC9340}{doi:10.17487/RFC9340}.
\BIBentrySTDinterwordspacing

\bibitem{q-fly}
D.~Sakuma \emph{et~al.}, ``An optical interconnect for modular quantum computers,'' \emph{arXiv preprint arXiv:2412.09299}, 2024.

\bibitem{science.1231298}
\BIBentryALTinterwordspacing
C.~Monroe and J.~Kim, ``Scaling the ion trap quantum processor,'' \emph{Science}, vol. 339, no. 6124, pp. 1164--1169, 2013, \href{https://dx.doi.org/10.1126/science.1231298}{doi:10.1126/science.1231298}.
\BIBentrySTDinterwordspacing

\bibitem{Lekitsch2017Blueprint}
\BIBentryALTinterwordspacing
B.~Lekitsch \emph{et~al.}, ``Blueprint for a microwave trapped ion quantum computer,'' \emph{Science Advances}, vol.~3, no.~2, p. e1601540, 2017, \href{https://dx.doi.org/10.1126/sciadv.1601540}{doi:10.1126/sciadv.1601540}.
\BIBentrySTDinterwordspacing

\bibitem{akhtar2023high}
\BIBentryALTinterwordspacing
M.~Akhtar \emph{et~al.}, ``A high-fidelity quantum matter-link between ion-trap microchip modules,'' \emph{Nature Communications}, vol.~14, no.~1, p. 531, Feb. 2023, \href{https://dx.doi.org/10.1038/s41467-022-35285-3}{doi:10.1038/s41467-022-35285-3}.
\BIBentrySTDinterwordspacing

\bibitem{PhysRevLett.130.050803}
\BIBentryALTinterwordspacing
V.~Krutyanskiy \emph{et~al.}, ``Entanglement of trapped-ion qubits separated by 230 meters,'' \emph{Phys. Rev. Lett.}, vol. 130, p. 050803, Feb 2023, \href{https://dx.doi.org/10.1103/PhysRevLett.130.050803}{doi:10.1103/PhysRevLett.130.050803}.
\BIBentrySTDinterwordspacing

\bibitem{NASA_1966}
{National Aeronautics and Space Administration}, ``Procedure for failure mode, effects, and criticality analysis (fmeca),'' \url{https://ntrs.nasa.gov/citations/19700076494}, 1966, nASA Technical Memorandum NASA-TM-X-65227, accessed via NASA Technical Reports Server.

\bibitem{sharma2018fmea}
K.~D. Sharma and S.~Srivastava, ``Failure mode and effect analysis (fmea) implementation: A literature review,'' \emph{Journal of Advance Research in Aeronautics and Space Science}, vol.~5, no. 1\&2, pp. 1--17, Apr. 2018, [Online]. Available: \url{https://www.researchgate.net/publication/333209894_Failure_Mode_and_Effect_Analysis_FMEA_Implementation_A_Literature_Review}.

\bibitem{Paul_Nobel}
\BIBentryALTinterwordspacing
W.~Paul, ``Electromagnetic traps for charged and neutral particles,'' Nobel Lecture, 1990.
\BIBentrySTDinterwordspacing

\bibitem{IonTrap_QC1}
\BIBentryALTinterwordspacing
J.~I. Cirac and P.~Zoller, ``Quantum computations with cold trapped ions,'' \emph{Phys. Rev. Lett.}, vol.~74, pp. 4091--4094, May 1995, \href{https://dx.doi.org/10.1103/PhysRevLett.74.4091}{doi:10.1103/PhysRevLett.74.4091}.
\BIBentrySTDinterwordspacing

\bibitem{qccd_proposal}
D.~Kielpinski, C.~Monroe, and D.~Wineland, ``Architecture for a large-scale ion-trap quantum computer,'' \emph{Nature}, vol. 417, pp. 709--11, 07 2002, \href{https://dx.doi.org/10.1038/nature00784}{doi:10.1038/nature00784}.

\bibitem{PhysRevLett.96.253003}
\BIBentryALTinterwordspacing
S.~Seidelin \emph{et~al.}, ``Microfabricated surface-electrode ion trap for scalable quantum information processing,'' \emph{Phys. Rev. Lett.}, vol.~96, p. 253003, Jun 2006, \href{https://dx.doi.org/10.1103/PhysRevLett.96.253003}{doi:10.1103/PhysRevLett.96.253003}.
\BIBentrySTDinterwordspacing

\bibitem{Benhelm_2008}
\BIBentryALTinterwordspacing
J.~Benhelm, G.~Kirchmair, C.~F. Roos, and R.~Blatt, ``Towards fault-tolerant quantum computing with trapped ions,'' \emph{Nature Physics}, vol.~4, no.~6, p. 463–466, Apr. 2008, \href{https://dx.doi.org/10.1038/nphys961}{doi:10.1038/nphys961}.
\BIBentrySTDinterwordspacing

\bibitem{RevModPhys.75.281}
\BIBentryALTinterwordspacing
D.~Leibfried, R.~Blatt, C.~Monroe, and D.~Wineland, ``Quantum dynamics of single trapped ions,'' \emph{Rev. Mod. Phys.}, vol.~75, pp. 281--324, Mar 2003, \href{https://dx.doi.org/10.1103/RevModPhys.75.281}{doi:10.1103/RevModPhys.75.281}.
\BIBentrySTDinterwordspacing

\bibitem{Eschner:03}
\BIBentryALTinterwordspacing
J.~Eschner, G.~Morigi, F.~Schmidt-Kaler, and R.~Blatt, ``Laser cooling of trapped ions,'' \emph{J. Opt. Soc. Am. B}, vol.~20, no.~5, pp. 1003--1015, May 2003, \href{https://dx.doi.org/10.1364/JOSAB.20.001003}{doi:10.1364/JOSAB.20.001003}.
\BIBentrySTDinterwordspacing

\bibitem{osada2021loading}
\BIBentryALTinterwordspacing
A.~Osada and A.~Noguchi, ``Deterministic loading of a single strontium ion into a surface electrode trap using pulsed laser ablation,'' 2021.
\BIBentrySTDinterwordspacing

\bibitem{Schioppo_2012}
\BIBentryALTinterwordspacing
M.~Schioppo \emph{et~al.}, ``A compact and efficient strontium oven for laser-cooling experiments,'' \emph{Review of Scientific Instruments}, vol.~83, no.~10, Oct. 2012, \href{https://dx.doi.org/10.1063/1.4756936}{doi:10.1063/1.4756936}.
\BIBentrySTDinterwordspacing

\end{thebibliography}
\end{document}